\documentclass[twocolumn,showpacs,aps,prb,superscriptaddress,floatfix]{revtex4}
\usepackage{graphicx}
\usepackage{dcolumn}
\usepackage{bm}
\begin{document}
\preprint{}
\newcommand{\toc}{TiOCl\,}
\newcommand{\tob}{TiOBr\,}
\newcommand{\tsco}{Ti$_{1-x}$Sc$_x$OCl\,}
\newcommand{\dscibo}{SrCu$_{2-x}$Mg$_x$($^{11}$BO$_3$)$_2$\,}


\affiliation{Department of Physics and Astronomy, McMaster University,
Hamilton, Ontario, L8S 4M1, Canada}
\affiliation{Canadian Institute for Advanced Research, 180 Dundas St. W.,
Toronto, Ontario, M5G 1Z8, Canada}

\author{J.P. Clancy}
\affiliation{Department of Physics and Astronomy, McMaster University,
Hamilton, Ontario, L8S 4M1, Canada}

\author{B.D. Gaulin}
\affiliation{Department of Physics and Astronomy, McMaster University,
Hamilton, Ontario, L8S 4M1, Canada}
\affiliation{Canadian Institute for Advanced Research, 180 Dundas St. W.,
Toronto, Ontario, M5G 1Z8, Canada}

\author{K.C. Rule}
\affiliation{Department of Physics and Astronomy, McMaster University,
Hamilton, Ontario, L8S 4M1, Canada}

\author{J.P. Castellan}
\affiliation{Department of Physics and Astronomy, McMaster University,
Hamilton, Ontario, L8S 4M1, Canada}

\author{F.C. Chou}
\affiliation{Center for Condensed Matter Sciences, National Taiwan University, Taipei 106, Taiwan.}

\title{Commensurate Fluctuations in the Pseudogap and Incommensurate spin-Peierls Phases of \toc}

\begin{abstract}
X-ray scattering measurements on single crystals of \toc reveal the presence of commensurate
dimerization peaks within both the incommensurate spin-Peierls phase and the so-called
pseudogap phase above T$_{C2}$. This scattering is relatively narrow in {\bf Q}-space indicating long correlation lengths exceeding
$\sim$ 100 $\AA$ below T$^*$$\sim$ 130 K.  It is also slightly shifted in {\bf Q} 
relative to that of the commensurate long range ordered state at the lowest temperatures, and 
it coexists with the incommensurate Bragg peaks below T$_{C2}$.  The integrated scattering over both 
commensurate and incommensurate positions evolves continuously with decreasing temperature for all temperatures 
below T$^*$$\sim$ 130 K. 
\end{abstract}

\maketitle

Low dimensional quantum magnets with singlet ground states are of intense current interest
due to the novelty of their ground states and their relationship to high temperature
superconductivity\cite{Review}.  Among quasi-one dimensional antiferromagnets, much attention has focused
on a small number of quasi-one-dimensional Heisenberg S=1/2 chain materials which undergo a spin-Peierls phase transition
to a dimerized singlet ground state\cite{spin-Peierls}.  CuGeO$_3$ was the first inorganic material to display a 
spin-Peierls phase transition of T$_C$$\sim$ 14 K\cite{CuGeO3}.  Its discovery allowed 
studies of this exotic state in large single crystal form and also in the presence of dopants\cite{CuGeO3doped}.  
These works substantially informed the discussion on the nature of this quantum ground state.  Earlier studies on organic 
spin-Peierls materials, such as TTF-CuBDT\cite{spin-Peierls} and MEM-(TCNQ)$_2$\cite{MEM} also attracted great interest, 
but were more limited due to the low density of magnetic moments that these materials possess.

Recently, quantum magnets based on Ti$^{3+}$ (3d$^1$) have been shown to display unconventional spin-Peierls 
states.  At low temperatures \toc and \tob both display dimerized magnetic singlet ground states.  On increasing
temperature both materials undergo discontinuous phase transitions at T$_{C1}$ to incommensurate states 
closely related to the dimerized ground state\cite{Seidel, tocNMR, Shaz, Krimmel, Lemmens, Sasaki}.  
On further raising the temperature, the 
incommensurate ordered phase undergoes an apparently continuous phase transition
at T$_{C2}$ to a disordered pseudogap phase. \toc displays an NMR signature within this pseudogap phase
between T$_{C2}$ and T$^*$$\sim$ 130 K, similar to that
observed in underdoped high temperature superconductors\cite{tocNMR}.  
T$_{C1}$ and T$_{C2}$ occur at $\sim$ 63 K and 93 K respectively in \toc\cite{Seidel, tocNMR, Krimmel, Ruckamp, Hemberger}, and at 
$\sim$ 27 K and 47 K in \tob\cite{Lemmens, Sasaki}.

The spin-Peierls states in \toc and \tob possess significant and interesting differences from previously studied spin-Peierls 
materials.  Their phase transition temperatures are higher by factors of between 2 to 4.  Furthermore, the singlet-triplet 
gap deduced from NMR\cite{tocNMR} and $\mu$SR measurements\cite{Baker} on \toc below T$_{C1}$ are $\sim$ 430 K and $\sim$ 420 K, respectively. 
These are high compared with other spin-Peierls systems and with its own T$_{C1,C2}$; 
${2E_g\over k_BT_{C1,C2}}$ $\sim$ 10 to 15, as opposed to the conventional BCS value of 3.5 
expected within mean field theory\cite{spin-Peierls}.

\toc crystallizes into the othorhombic FeOCl structure, and is comprised of Ti-O bilayers separated by chlorine ions along 
the c-direction\cite{crys, Seidel}. It displays room temperature lattice parameters of a=3.79 $\AA$, b=3.38 $\AA$ and 
c=8.03 $\AA$\cite{crys}.  At 
temperatures of 200 K and above its magnetic susceptibility, $\chi$(T), can be described by an S=1/2 Heisenberg chain
model, with a near-neighbor exchange constant J=660 K\cite{Seidel}.  
The implied reduction of the magnetic dimensionality of the 
Ti-O bilayers, from two to one, has been proposed as 
originating from ordering of Ti$^{3+}$ orbital degrees of freedom\cite{Seidel}. 

In this Letter we report x-ray scattering measurements on \toc which show commensurate fluctuations of a 
dimerized structure that first appear at T$^*$$\sim$ 130 K, the onset of the pseudogap phase. 
 We further show that the intensity and width of the scattering from these fluctuations peak at T$_{C2}$, 
but the commensurate fluctuations coexist with the 
incommensurate ordered structure for all temperatures above T$_{C1}$.  Finally we show that the growth of the Q-integrated 
dimerization scattering is continuous at all temperatures measured.

\begin{figure}[b]
\includegraphics{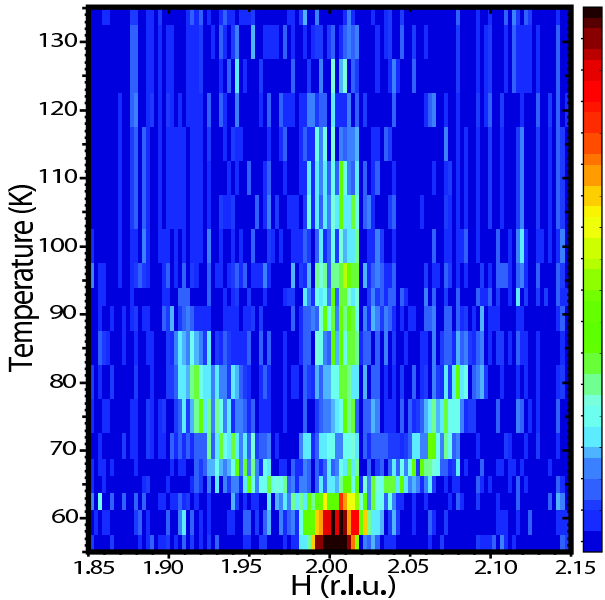}
\caption{(Color) A color contour map of x-ray scattering intensity as a 
function of temperature in \toc.  H-scans of the form (H, 1.5, 1) are shown with background scans at 200 K subtracted.  
One clearly observes commensurate dimerization scattering at (2, 1.5, 1) in both the incommensurate (T$_{C1}$$\sim$63 K to 
T$_{C2}$$\sim$93 K) and pseudogap phases (T$_{C2}$ to T$^*$$\sim$130 K). }
\end{figure}

Single crystal samples of \toc were grown by the chemical vapor transport method, as reported earlier\cite{Seidel}.  
The crystal studied had approximate dimensions of 2.0 by 2.0 by 0.1 mm.  
The sample was mounted on the cold finger of a closed cycle refrigerator and aligned within a Huber four circle goniometer.  
Cu-K$\alpha$ ($\lambda$ = 1.54 A) radiation from a rotating anode source was employed in all measurements reported.  
The temperature of the sample was stabilized to $\sim$$\pm$ 0.01 K.  

X-ray scattering scans of the form (H, 1.5, 1), (2, K, 1) and (2, 1.5, L) were performed around 
the commensurate dimerization wavevector (2, 1.5, 1) and several other equivalent ordering wavevectors 
at low temperatures.  We ultimately focused around the (2, 1.5, 1) dimerization wavevector, as the
intensity of the superlattice Bragg peaks was strongest there.  
Figs.\ 1 and 2 show the H and K scans which were carried out as a function of temperature.  
Fig.\ 1 shows a color contour map containing all of the 
H scans performed between 55 K and 150 K, and the main qualitative features of this letter can be found in it.  
Fig.\ 2 shows representative H and K scans at three temperatures - in the commensurate spin-Peierls phase at 55 K, 
the incommensurate phase at 75 K/70 K, and the pseudogap phase at 105 K.  

All data sets have had a background data set, collected at 200 K, subtracted from them.  This is necessary as 
weak $\lambda$/2 contamination of the incident x-ray beam produces Bragg scattering at the commensurate, (2, 1.5, 1) position
in reciprocal space.  This $\lambda$/2 scattering, whose strength is comparable to the $\lambda$ scattering in 
the pseudogap phase ($\sim$ 0.1 counts/sec), is temperature independent and can easily be subtracted off to reveal a background-free signal.

Several features are immediately clear from examination of Fig.\ 1.  The previously observed incommensurate 
Bragg peaks\cite{Krimmel} which 
appear continuously below T$_{C2}$$\sim$ 93 K, evolve towards the commensurate dimerization wavevector with decreasing 
temperature, giving rise to the transition to the commensurate, dimerized spin-Peierls state below T$_{C1}$$\sim$ 63 K.
However, most importantly, we observe commensurate scattering centered very near (2, 1.5, 1), which coexists with the
incommensurate spin-Peierls state, and can be observed to temperatures as high as T$^*$$\sim$ 130 K.

\begin{figure}
\includegraphics{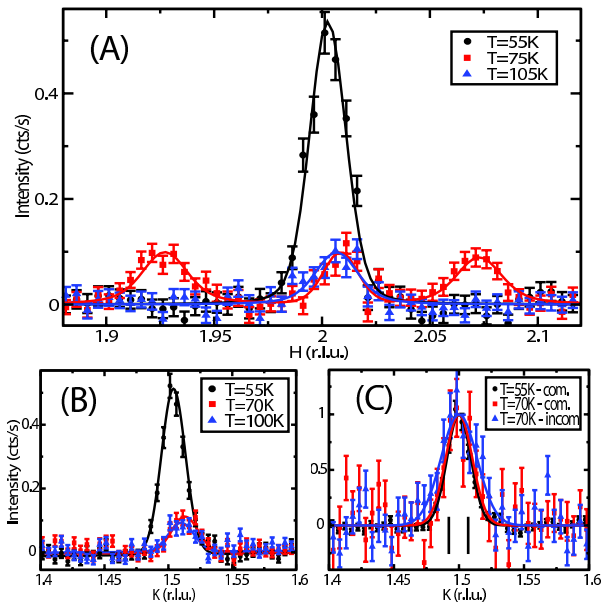}
\caption{(Color) (a) Representative H-scans of the form (H, 1.5, 1).  (b) Representative K-scans of the
form (2, K, 1).  (c) K-scans shown in (b) are scaled so that peak intensities 
are the same, and are centered such that peaks occur at K=1.5.  The incommensurate
scattering at T=70 K is broader in K than either the commensurate scattering above or below T$_{C1}$.
This is attributed to weak incommensuration along K, such that the incommensurate scattering
occurs at (2$\pm\delta$, 1.5$\pm\epsilon$, 1).  The vertical lines indicate the expected position of the 
incommensurate peaks in K for T=70 K.}
\end{figure}

This commensurate dimerization scattering above T$_{C1}$ has not been observed previously in either \toc or \tob, and 
it possesses several very interesting characteristics.  Close examination of the detailed H and K scans shown in Fig.\ 2
reveals that this commensurate scattering is peaked close to, but not precisely at, the commensurate wavevector characterizing
the low temperature dimerized state below T$_{C1}$.  Figs.\ 2a and 2b show this scattering peaks up at $\sim$
(2.005, 1.505, 1) rather than at (2, 1.5, 1).  The incommensurate scattering we observe between T$_{C1}$ and T$_{C2}$ is also peaked up at 
(2$\pm\delta$, 1.5, 1); that is it is centered on K=1.5, like the low temperature commensurate scattering.  
Fig.\ 2c shows scaled and centered K-scans of the commensurate scattering in the 
low temperature phase at 55 K, as well as commensurate and incommensurate scattering between T$_{C1}$ and T$_{C2}$ at 70 K.  
We see that the incommensurate scattering at 70 K is 
broader in K than either the low temperature commensurate scattering or the commensurate scattering above T$_{C1}$.  
This is explained and indeed expected as a consequence of the known\cite{Krimmel}, weak incommensuration along K, such that the 
precise incommensurate wavevector is (2$\pm\delta$, 1.5$\pm\epsilon$, 1) with $\delta\sim$0.06 and $\epsilon\sim$0.006 for \toc at 
70 K as shown in Figs.\ 2b and 2c\cite{Krimmel}.  The incommensurate phase in \tob is characterized by a similar form of the 
incommensurate wavevector - with incommensuration in both H and K\cite{Sasaki}.  Our measurements
see this as a broadening in wavevector K, rather than as distinct incommensurate peaks, due to the relatively low resolution 
of our measurements which integrate over this very weak $\epsilon$-incommensuration.

The new commensurate scattering above T$_{C1}$ in \toc is relatively sharp in {\bf Q}-space at 
all temperatures measured, indicating correlation lengths in excess of $\sim$ 100 $\AA$ even within the pseudogap phase.  
This commensurate scattering is therefore distinct from conventional critical scattering (as measured around 
the {\it incommensurate} Bragg positions at and above T$_{C2}$ in \tob\cite{Sasaki}) wherein the correlation lengths 
diverge at the phase transition, typically as power laws.  However, critical fluctuations are usually small at (T-T$_C$)/T$_C$ $\sim$ 0.1, 
which corresponds to 9 K above a phase transition with T$_C$$\sim$ 90 K.  Fig.\ 3 shows $\Gamma$, the wavevector-H width of the new 
commensurate scattering, as well as of the incommensurate scattering, as a function of temperature.
These widths were extracted from fits of the H-scans, shown in Figs.\ 1 and 2, to a Lorentzian form for the scattering. 
The main panel shows fits 
to the raw data itself, while the inset shows the intrinsic width of the Lorentzian describing the scattering, convoluted in one dimension 
by the experimental resolution function appropriate to the measurement.  This result shows that the 
new commensurate scattering above T$_{C1}$ has a narrow, but non-resolution limited lineshape for all temperatures above
T$_{C1}$.  The width of the scattering, and hence the inverse correlation length, peak near T$_{C2}$.  The same trend is seen
when a resolution convolution of the scattering is explicitly included to extract quantitative estimates of the correlation
lengths.  We see that typical correlation lengths away from T$_{C2}$ are $\xi_a$ $\sim$ 1/HWHM $\sim$ 
1/(${2\pi \over a}$ $\times$ 0.002) $\sim$ 300 $\AA$, dropping to $\sim$ 100 $\AA$ in the vicinity of T$_{C2}$. 

We therefore conclude that the new commensurate scattering above T$_{C1}$ is due to fluctuations, rather than 
long range order (LRO).  This is consistent with NMR measurements\cite{tocNMR} which show static LRO to occur only below 
T$_{C2}$.  In addition, while several experimental techniques show evidence for the pseudogap phase below T$^*$, 
bulk characterization by susceptibility\cite{Seidel, Ruckamp} and heat capacity\cite{Ruckamp, Hemberger} 
provide no evidence of an additional phase transition above T$_{C2}$.    
Fig.\ 3 also shows the analysis for $\Gamma$, the H-width of the incommensurate scattering between 
T$_{C1}$ and T$_{C2}$.  This indicates that the scattering is broader than resolution-limited, on the basis of comparison 
with the commensurate LRO scattering at T=55 K.  The interpretation of the width of the incommensurate scattering
is complicated by the weak $\epsilon$-incommensuration
along K, which may induce a finite H-width to the measured scattering by virtue of our resolution.  
However, the most straightforward interpretation of these results is that the incommensurate scattering also displays a 
long but non-infinite correlation length $\sim$ 200 $\AA$, because coexisting commensurate fluctuations and 
incommensurate order compete against each other between T$_{C1}$ and T$_{C2}$.  Examination of their 
temperature dependencies also supports such an interpretation.

\begin{figure}[t]
\includegraphics{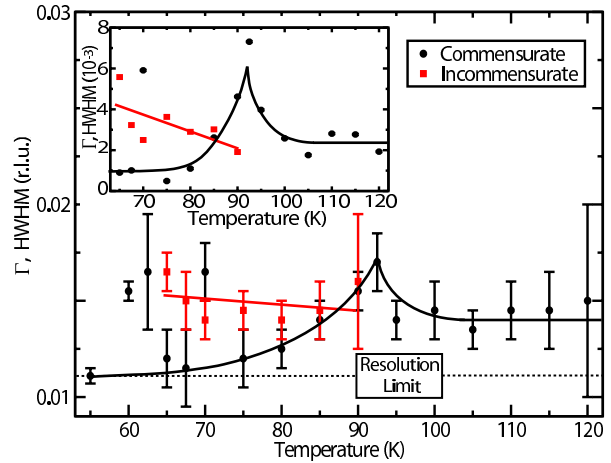}
\caption{(Color online) $\Gamma$, the H-width of the commensurate scattering above T$_{C1}$, as well as the incommensurate scattering between
T$_{C1}$ and T$_{C2}$ is shown.  The main panel shows widths extracted from Lorentzian fits to the raw data, while the 
inset shows intrinsic Lorentzian widths extracted from 1D resolution-convoluted fits to the data.  All lines provided are guides-to-the-eye.}
\end{figure}

The temperature dependence of the commensurate and incommensurate dimerization scattering is shown in Fig.\ 4.  This 
integrated intensity is obtained by performing integrations over a small range of wavevectors around either the incommensurate
or commensurate wavevectors such that all of the relevant scattering is captured.
One clearly sees the continuous rise of the incommensurate intensity near T$_{C2}$$\sim$ 93 K, and the discontinuous
rise in the commensurate scattering near T$_{C1}$$\sim$ 63 K.  Most importantly, one sees the continuous rise of the
commensurate fluctuation scattering near T$^{*}$$\sim$ 130 K, which peaks at T$_{C2}$ and decreases slowly until T$_{C1}$.

We conclude that the pseudogap phase between T$^*$ and T$_{C2}$ is characterized by the growth of
commensurate dimerization fluctuations with correlation lengths of many unit cells. Below T$_{C2}$ incommensurate
Bragg-like scattering co-exists with the commensurate fluctuations.  The concomitant rise of the incommensurate 
scattering intensity along with the peak in the temperature dependence of the 
integrated intensity of the commensurate fluctuations at T$_{C2}$ suggests
a competition between these two states in the regime between T$_{C1}$ and T$_{C2}$.

Related phenomena may occur in Mg-doped CuGeO$_3$\cite{Wang}, wherein a spin gap is established at a higher temperature 
than that which characterizes the spin-Peierls LRO.  This pseudo-gap regime displays commensurate
dimerization scattering, with long, but not infinite, correlation lengths, and it has been attributed to impurity-induced interchain 
interactions. 

\begin{figure} [t]
\includegraphics{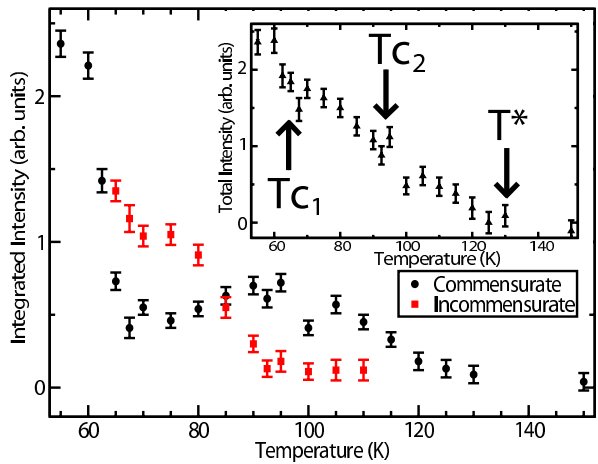}
\caption{(Color online) The integrated x-ray scattering around the commensurate position (2, 1.5, 1) is shown as a function
of temperature with the black circles.  The red squares show the integrated incommensurate scattering
around (H$\pm$$\delta$, 1.5, 1).  The inset shows the integrated scattering over all H in scans of the form 
(H, 1.5, 1), as a function of temperature.  The growth of the Q-integrated scattering is continuous 
for all temperatures below T$^*$$\sim$ 130 K.}
\end{figure}

It is also straightforward to integrate up all the scattering in scans of the form shown in Figs.\ 1 and 2, covering both
the commensurate and incommensurate wavevectors.  This temperature dependence is shown in the inset to Fig.\ 4.  Remarkably 
this integrated intensity varies smoothly with temperature from T$^*$, with no sign of a discontinuity at T$_{C1}$.  The
temperature dependence of the intensity of the commensurate LRO scattering below T$_{C1}$ is known to be almost temperature 
independent\cite{Krimmel}.  Hence, we see a linear growth in the overall dimerization scattering below T$^*$, 
saturating into a commensurate LRO spin-Peierls state below T$_{C1}$. 

Most phase transitions are characterized by an order parameter and order parameter fluctuations, 
both of which possess the same wavevector.  However there are examples of materials whose fluctuations above
a phase transition occur at a different wavevector than that which characterizes the LRO state below T$_C$.
UNi$_2$Al$_3$, for example, exhibits an incommensurate magnetically ordered state below T$_N$$\sim$ 4.6 K\cite{Shroeder} 
which is preceded by dynamic short range fluctuations at the commensurate magnetic 
wavevector\cite{Gaulin} which characterizes the ordered state in its sister compound UPd$_2$Al$_3$\cite{UPA}.  
This is believed to arise due to competition between RKKY interactions and Kondo
screening in this heavy fermion metal.  Competing interactions leading to frustration are widely discussed in
the context of the incommensurate spin-Peierls phases of both \toc and \tob.  Ruckamp et al\cite{Ruckamp}, 
for example, have proposed that the incommensurate order in \toc originates from competing in-phase and 
out-of-phase arrangements of dimers within the Ti-O layers which make up the bilayers.  
Our data suggests that such competing interactions not only give rise to the incommensurate spin-Peierls 
state, but also lead to contending yet co-existing ground states between T$_{C1}$ and T$_{C2}$.  

To conclude, we have observed commensurate dimerization scattering which first appears at the onset of
the pseudogap phase, T$^*$$\sim$ 130 K, and which coexists with the incommensurate state below T$_{C2}$.  
This commensurate scattering exhibits long, but finite correlation lengths at all temperatures measured, and both 
its integrated intensity and inverse correlation length peak at T$_{C2}$.  Although similar commensurate scattering 
has yet to be observed in TiOBr, the combination of competing interactions and incommensurate 
structure which is common to both materials could easily give rise to an analogous effect.

We wish to acknowledge very helpful discussions with T. Imai. This work was supported by NSERC of Canada and NSC of Taiwan.
%
%
%
%
%
%
%
%
%
%

\end{document}